\documentclass[aps,prstab,twocolumn,showpacs,superscriptaddress,groupedaddress]{revtex4}  
\usepackage{graphicx}  
\usepackage{dcolumn}   
\usepackage{natbib}
\usepackage[caption=false]{subfig}
\usepackage{bm}        
\usepackage{amssymb}   
\begin{document}



\title{High Efficiency, Multi-Terawatt X-ray free electron lasers}
%
\affiliation{University of California, Los Angeles, California 90095, USA}
\affiliation{Stanford Linear Accelerator Center, Menlo Park, California 94025, USA}
\author{C.~Emma} \affiliation{University of California, Los Angeles, California 90095, USA}
\author{K.~Fang} \affiliation{Stanford Linear Accelerator Center, Menlo Park, California 94025, USA}
\author{J.~Wu} \affiliation{Stanford Linear Accelerator Center, Menlo Park, California 94025, USA}
\author{C.~Pellegrini} \affiliation{University of California, Los Angeles, California 90095, USA}\affiliation{Stanford Linear Accelerator Center, Menlo Park, California 94025, USA}
%
%
%
\vskip 0.25cm
\date{\today}

\begin{abstract}
We study high efficiency, multi-terawatt peak power, few angstrom wavelength, X-ray Free Electron Lasers
(X-ray FELs). To obtain these characteristics we consider an optimized undulator design: superconducting,
helical, with short period and built-in strong focusing. This design reduces the length of the breaks between
modules, decreasing diffraction effects, and allows using a stronger transverse electron focusing. Both effects
reduce the gain length and the overall undulator length. The peak power and efficiency depend on the
transverse electron beam distribution and on time dependent effects, like synchrotron sideband growth.
The last effect is identified as the main cause for reduction of electron beam microbunching and FEL peak power.
We show that the optimal functional form for the undulator magnetic field tapering profile, yielding the
maximum output power, depends significantly on these effects. The output power
achieved when neglecting time dependent effects for an LCLS-like X-ray FEL with a 100 m long tapered
undulator is 7.3 TW, a 14 $\%$ electron beam energy extraction efficiency. When these effects are included the
highest peak power is achieved reducing the tapering rate, thus minimizing the reduction in electron
micro-bunching due to synchrotron sideband growth. The maximum efficiency obtained for this case is 9
$\%$, corresponding to 4.7 TW peak radiation power. Possible methods to suppress the synchrotron sidebands, and further enhance the FEL peak power, up to about 6 TW by increasing the seed power, are discussed.
\end{abstract}

\maketitle

\section{Introduction}

Maximising the extraction efficiency of a Free Electron Laser (FEL) via undulator tapering has been examined theoretically \cite{KMR} \cite{Fawley2002537}, numerically \cite{PhysRevSTAB.Y.Jiao} \cite{SY} \cite{Mak} and demonstrated experimentally from the microwave \cite{LANLtaper} down to hard X-ray wavelengths \cite{Ratnertapering}. At hard X-ray wavelengths the achieved efficiencies are a factor of 3 larger than the exponential saturation value \cite{Ratnertapering}. Single molecule imaging applications require TW-level X-ray pulse power and thus an improvement in the extraction efficiency of more than an order of magnitude in the next generation of X-ray FELs \cite{SingleMoleculeImaging}.
This has sparked renewed interest in the community with a number of recent studies specifically devoted to finding the optimal tapering law using a model based approach and a form of parametric optimization \cite{SY} \cite{Mak} \cite{DurisTESSA}. With the exception of Ref. \cite{DurisTESSA}, which considers longer wavelengths than we do in this paper, the optimizations presented have dealt exclusively with the time independent physics of tapered FELs, producing a taper profile which maximises the extraction efficiency ignoring time dependent effects like the synchrotron sideband instability \cite{KMR}. The physical difference between time independent and time dependent optimization arises from noise in the electron beam current distribution and slippage of the radiation field. This can drive the amplification of parasitic frequencies and the sideband instability, causing temporal fluctuations in the electric field profile, particle detrapping and eventually taper saturation. 

In this paper we show that the solution obtained for the optimal taper profile in time independent simulations does not yield the maximum extraction efficiency when fully time dependent physics is included in the dynamics of the the electron beam and radiation field system. We study the optimization problem by following the multidimensional scan method of Ref.  \cite{PhysRevSTAB.Y.Jiao} for a superconducting, 2 cm period helical undulator with built in focusing. This undulator design is optimized for maximum efficiency, reduction of intra module undulator length, strong transverse focusing, short gain length and minimum total undulator length. The characteristics are given in Table 1. The simulations are performed for a transversely flat electron beam distribution as this maximises the output power, as described in Ref. \cite{EMMACPRSTAB}.

The paper is divided as follows. In section II we discuss the undulator design, outlining the undulator parameters and the feasibility of practical realization. In section III we discuss the self-seeding method and simulate the SASE FEL upstream of the optimized tapered undulator. In sections IVa and IVb we present the time independent and time dependent tapering optimization results and discuss the differences between the two. In section V we analyze in detail the sideband instability, its onset, amplification and its impact on particle dynamics and taper saturation. Finally we discuss methods to suppress the instability and outline future work on high efficiency tapered undulators and optimization algorithms.

\begin{figure}
\includegraphics[scale=0.4]{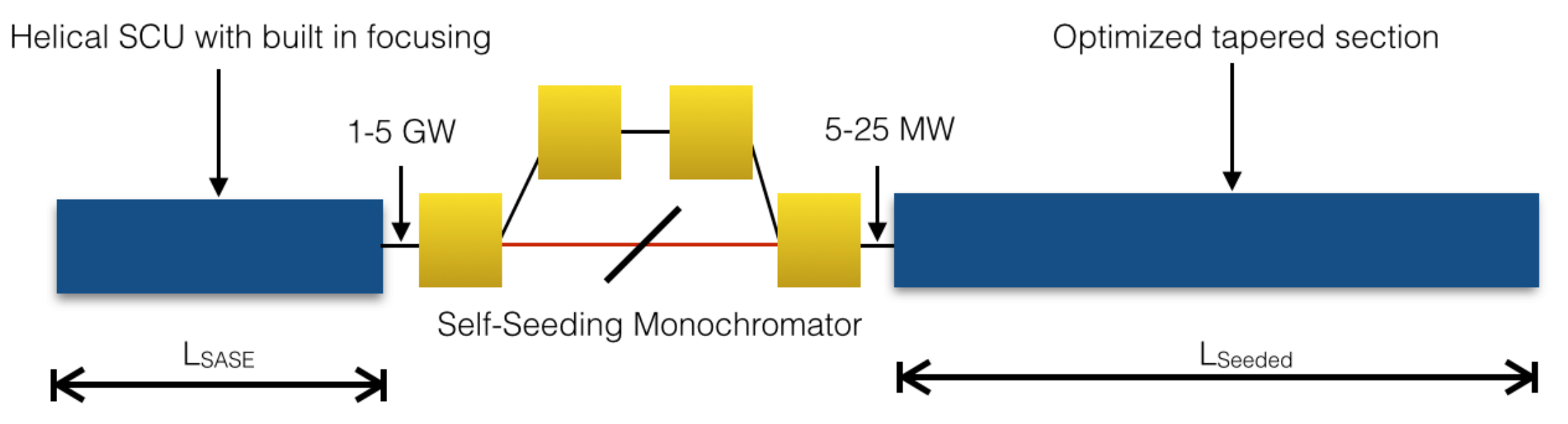}
\caption{Schematic of the undulator for hard X-ray multi TW peak power output, designed to achieve high extraction efficiency in the shortest possible distance.}
\end{figure}

\section{Undulator Design}

We apply the tapering optimization method \cite{PhysRevSTAB.Y.Jiao} to an undulator designed specifically to achiveve TW power X-ray pulses in the shortest possible undulator length. Our ideal undulator is superconducting, with a short 2 cm period and a peak on axis field $B_0$ of 1.6 T. For a double helix bifilar magnet with equal and opposite currents this field is given by \cite{Smythe}: 

\begin{equation}
B_0=\frac{4k_uI}{10^5}\left[k_u a K_0(k_u a)+K_1(k_u a)\right]
\end{equation}

where $I$ is the current in the coils, $k_u=2\pi/\lambda_u$ is the undulator wavenumber, $a$ is the helix radius and $K_0$ and $K_1$ are modified Bessel functions. For a helical bore radius $a$=7.5 mm the total current required through the coils is $I=$ 484 A which, considering coils of $\sim$ mm$^2$ surface area, gives a current density below the critical value for superconducting NbTi or Nb3Sn wires. From the point of view of operation a superconducting undulator has advantages such as resistance to radiation damage and reduced sensitivity to wakefields, for a more detailed description see Ref. \cite{PaulEmmaSCU}. The undulator is helically polarized as this increases the effect of refractive guiding in the post-saturation regime and improves the FEL performance \cite{PhysRevA.24.1436}. 

In order to accomodate diagnostics a realistic undulator design must include periodic break sections, with longer breaks adversely affecting performance. This is due essentially to three effects. Firstly, diffraction effects are critical to the performance of a tapered FEL particularly for long, multiple Rayleigh length undulators. While these effects are mitigated by refractive guiding inside the undulator, there is no guiding during the break sections and the radiation size increases exponentially, reducing the field amplitude, causing particle detrapping and limiting the extraction efficiency. Secondly, a break of length $L_b$ introduces a phase error $\Delta \Psi \sim L_b\delta/\gamma^2=2n\lambda_r\eta$ for a particle with relative energy offset $\eta=\delta\gamma/\gamma_r$ with respect to the resonant particle. Thus longer break sections increase electron phase mixing and reduce the bunching factor. Finally as a practical consideration, for a given total undulator length, longer break sections reduce the length of magnetic elements limiting the electron deceleration and over-all extraction efficiency. 
To minimize the break length we superimpose the focusing quadrupole field on the helical undulator field, similar to the design successfully tested in Ref. \cite{VISA}. One advantage of distributed quadrupole focusing is the possibility to operate at small betatron beta function, due to the reduced FODO lattice length $L_f$. This minimizes the transverse beam envelope oscillation $\Delta \beta^2/\beta_{av}^2=\beta_{av} L_f/(\beta_{av}^2-L_f^2)$ which also degrades the FEL performance \cite{SvenNara2012}. In our study the undulator magnetic field is tapered continuously and the section length is chosen to be 1 m, close to the 3-D gain length with 20 cm breaks in between. 

Although this kind of undulator has never been constructed in the past, the parameters presented in this design are similar to what is currently being considered for an LCLS-II-like planar superconducting undulator with the addition of built in quadrupole focusing \cite{PaulEmmaSCU}. A full engineering and tolerance study of this undulator is needed before we can be confident that it is a feasible option for future high efficiency X-ray FEL facilities. 

\begin{table}[t]
\caption{\label{tab:table1}GENESIS Simulation Parameters}
\begin{ruledtabular}
\begin{tabular}{lr}
Parameter Name&Parameter Value\\
\hline\hline
Beam Energy  & 12.975 GeV \\
Peak Current  & 4000 A \\
Normalized Emittances & 0.3/0.3 $\mu$ m rad \\
Average beta function & 5 m\\
RMS Energy Spread  & $10^{-4}$\\
Bunch Length &6 fs\\
\\
Seed radiation power & 5-25 MW \\
Radiation Wavelength  & 1.5 $\AA$\\
Rayleigh Length & 10 m\\
\\
Undulator Period & 2 cm\\
Undulator Parameter & 3\\
Quadrupole Focusing Strength &26.4 T/m\\
Undulator Section Length & 1 m\\
Undulator Break Length & 20 cm\\
\\
FEL parameter & 1.66 $\times 10^{-3}$ $ $\\
3-D Gain Length & 65 $  $cm\\
\end{tabular}
\end{ruledtabular}
\end{table}

\begin{figure}
\includegraphics[scale=0.45]{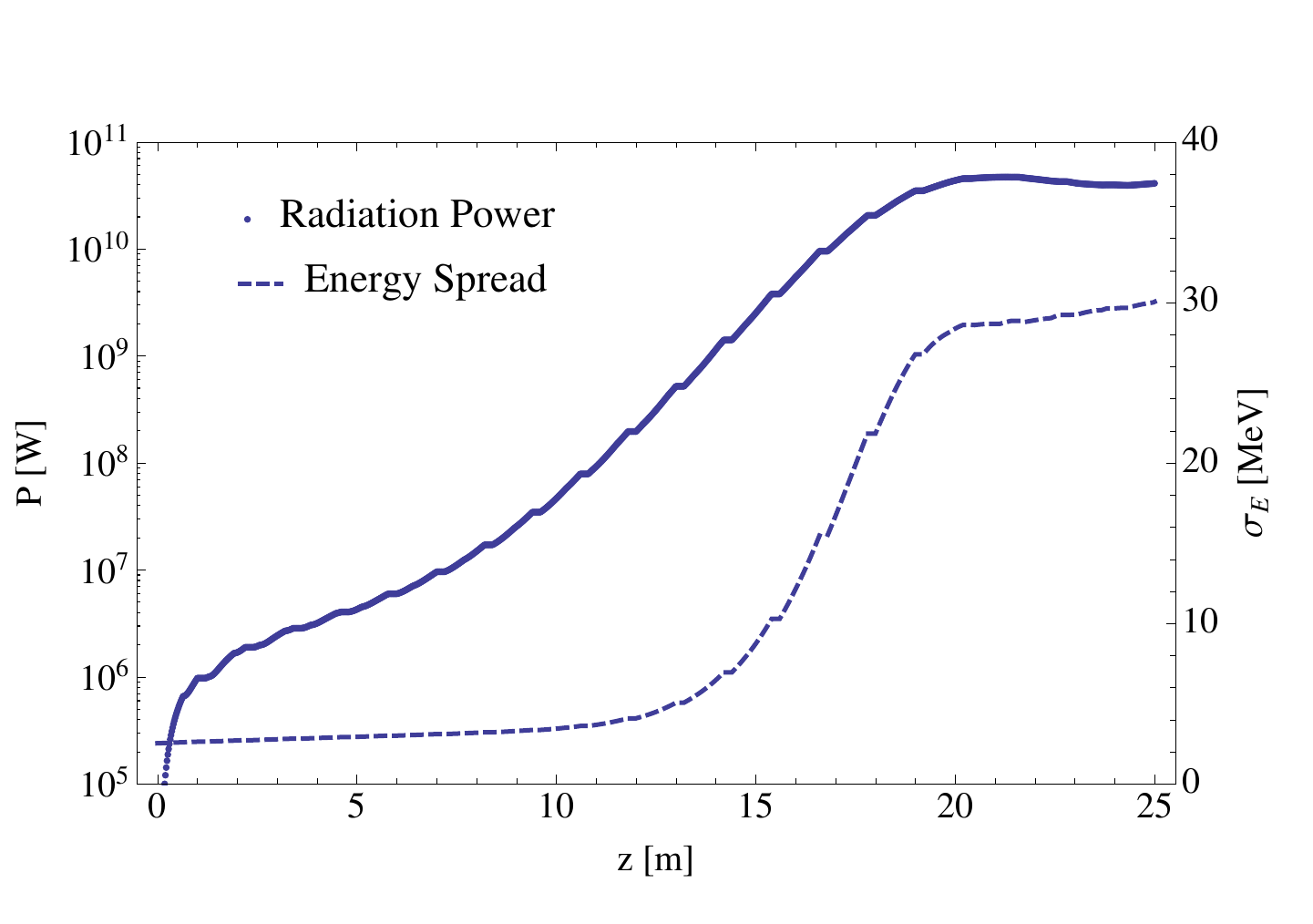}
\caption {SASE power and energy spread upstream of the self-seeding chicane. The energy spread at 1 GW of power is 3.1 MeV.}
\end{figure}

\begin{figure*}
\includegraphics[scale=0.45]{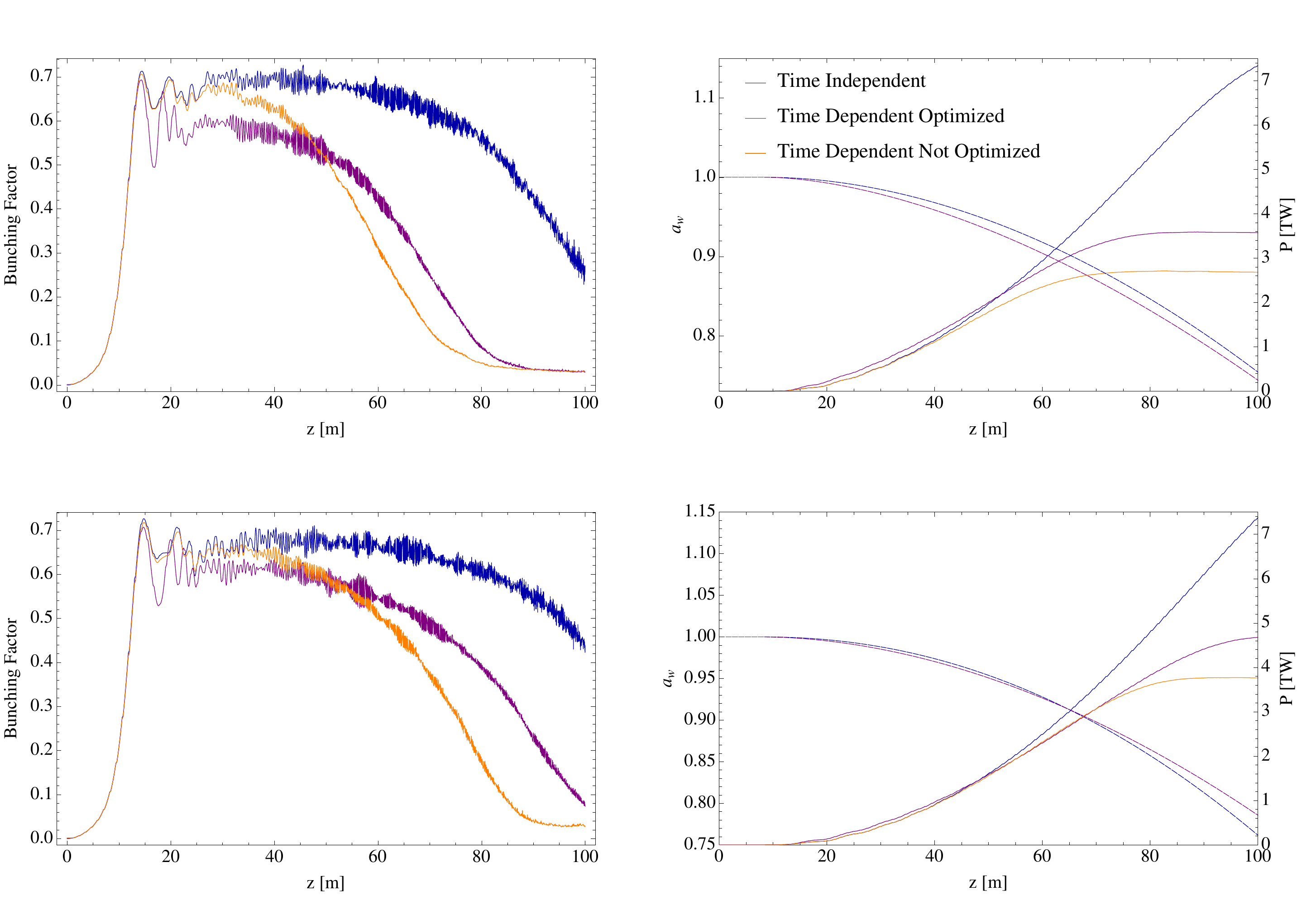}
\caption {Bunching factor (left), taper profiles and FEL radiation power evolution (right) obtained from time independent and time dependent optimization. The top plots correspond to an input energy spread $\sigma_{E,0}$=3.1 MeV consistent with the SASE result from Fig. 1. The bottom plots are an alternate case with $\sigma_{E,0}$=1.5 MeV. In both cases z=0 is after the self-seeding monochromator and the input seed power is 5 MW.}
\end{figure*}

\section{Self Seeding Stage}

We consider generating a monochromatic seed pulse using self-seeding design based on a single crystal monochromator similar to what is currently installed at LCLS \cite{AmmanSelfSeeding}. We assume that the tapered section follows a self-seeding chicane which completely eliminates the beam microbunching, and is preceded by a SASE section as shown in Fig.1. The energy spread at the start of the tapered section is then determined by the SASE FEL process, spontaneous emission losses in the SASE section \cite{SaldinSpontaneous} and the laser heater induced energy spread set to suppress the microbunching instability \cite{LaserHeater}. We perform a simulation of the SASE section assuming an initial RMS energy spread due to the linac of $10^{-4}$, and the results are shown in Fig. 2.  

For effective seeding the radiation power must exceed the electron beam shot noise power \cite{GiannessiTrieste}:

\begin{equation}
P_{noise}\approx \gamma mc^2\omega_r\rho^2/2
\end{equation}

by a wide margin. For our parameters this evaluates to 36 kW. We therefore decide to start the seeded section with 5 MW of power which requires 1 GW of SASE power incident on the monochromator. This sets the length of the SASE section $L_{SASE}=$ 13.4 m and the input energy spread $\sigma_E=3.1$ MeV. As pointed out in Ref. \cite{AmmanSelfSeeding}, in practice there is a trade-off between seed power and energy spread at the start of the seeded section. Here we study the two cases of 5 MW and 25MW seed power leaving a detailed analysis of how this trade-off impacts the tapered FEL performance for future work.

\section{Tapering Optimization}
\subsection{Time Independent}

We first obtain the optimal taper profile, maximizing the output power for a fixed 100 m undulator length in time independent simulations using the three dimensional FEL particle code GENESIS \cite{Reiche1999243}. The tapering law is written as:

\begin{equation}
a_w(z)=a_{w0}\times\left(1-c\times(z-z_0)^d\right)
\end{equation}

where the parameters $z_0,c,d$ are obtained by mutlidimensional scans which maximise the output power. The quadrupole focusing can also be tapered to further increase the extraction efficiency as shown in Ref. \cite{PhysRevSTAB.Y.Jiao} but that will not be considered in this study. The optimal taper profile obtained from time independent optimization is shown in Fig. 3. The tapering order is approximately quadratic, which follows qualitatively from the fact that in time independent simulations the bunching factor and trapping fraction remain nearly constant in the tapered section and the dominant radiative process is coherent emission. The peak output power is 7.3 TW with an extraction efficiency of 14 $\%$. It is important to note that there is no sign of the taper power saturating in the time independent case, which is not the case when time dependent effects are included. 

\begin{figure}[t]
\includegraphics[scale=0.35]{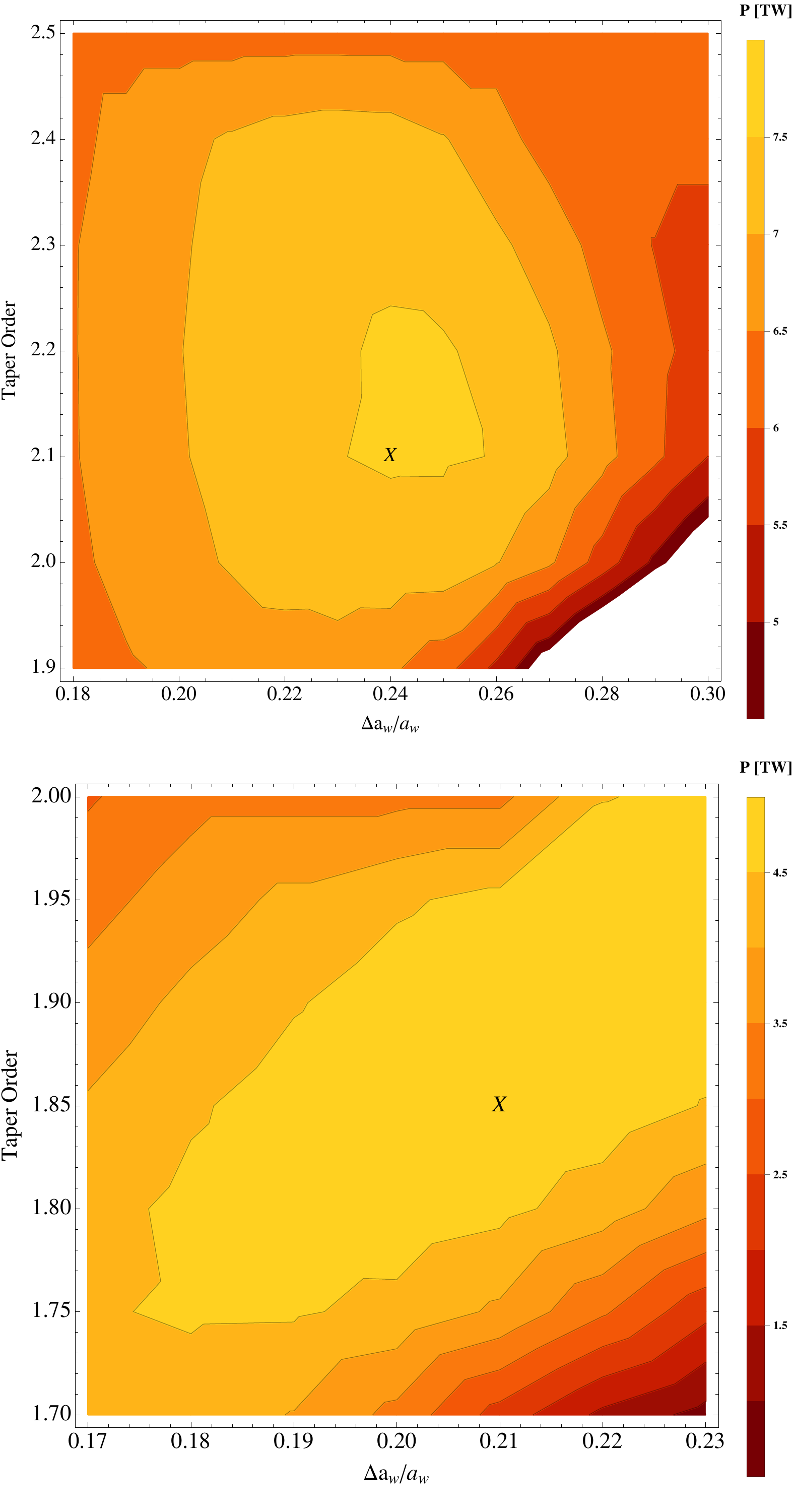}
\caption{Maximum radiation power as a function of taper order d and taper amplitude $\zeta$ for time independent (top) and time dependent (bottom) simulations for the case of 1.5 MeV input energy spread.}
\end{figure}

\subsection{Time Dependent Optimization}

Using the optimal taper starting point obtained from time independent simulations, we perform time dependent scans over the taper order $d$ and the taper strength $\Delta a_w/a_w = c\times (L_w-z_0)^d$. As shown in Fig. 3, the values of the taper order and taper strength yielding the maximum power in time independent simulations are not the optimal choice of parameters once time dependent effects are included. The variation in peak power is more sensitive to variations in the taper profile in the time dependent cases. The optimal taper order is weaker than quadratic, and is reduced compared to the time independent case. This is due to the FEL increased sensitivity to particle detrapping when electron beam shot noise and multiple frequency effects are included. 

Since the coherent emission power is proportional to the product of the number of trapped particles and the change in resonant energy (taper strength), a slower taper preserves the trapping for longer, maximising the product and the over-all extraction efficiency. In the large energy spread case it is important to note that the time dependent optimized taper profile has a slower taper order but a larger over-all deceleration rate. This results in a worse particle capture in the early stages of the tapered section (z=10-50m) but a reduction in detrapping in the remainder of the undulator. This can be understood by examining the functional form for the resonant phase:

\begin{equation}
\sin\Psi_R(z)= \chi\frac{|a_w'(z)|}{E(z)}
\end{equation}

where $\chi=(2*me*c^2/e)(\lambda_w/2\lambda_s)(1/\sqrt{2}{[JJ]})$ is a constant independent of $z$ and $E(z)$ is the electric field amplitude. The time dependent optimized taper reduces $|a_w'(z)|$ in the second half of the undulator z=50-100m, maintaining a larger bucket area in the region where the amplitude of the sidebands is more appreciable and the system is more sensitive to detrapping. From this is clear from this that a fully optimized form of the taper profile should have an improved capture rate in the early stages with a profile similar to what one obtains from time independent optimization, and a slower decrease in the undulator field in the later stages when time dependent effects are more appreciable. This requires a more elaborate functional form for $a_w(z)$ and will be investigated in future work. 

\subsection{Effect of the energy spread}

The input energy spread is a critical parameter for the performance of a tapered X-ray FEL. We study this by performing the same time dependent optimization for two cases both starting with a 5MW seed: the self-seeded case with an energy spread of 3.1 MeV and an alternate case with 1.5 MeV. In practice the alternate case could be achieved by considering a double-bunch system, where two closely spaced bunches are separated before the entrance to the undulator, the first bunch is sent through the undulator to lase producing the seed radiation and is discarded prior to the self-seeding chicane. The trailing bunch is transported outside of the undulator and is recombined with the seed pulse downstream of the self-seeding chicane. In this scheme the seeded bunch would have an RMS energy spread set only by the linac and the laser heater, around 1.5 MeV for our beam parameters.

As is evidenced in Fig. 3. the low energy spread case achieves a higher peak power, 4.7 TW compared to 3.7 TW after the time dependent optimization. The taper saturation is also delayed due to a decrease in sideband induced particle detrapping. In both cases electron emission into the lower synchrotron sideband mode causes detrapping from the high energy region of the stable phase space area. Furthermore, scattering of the electrons from interaction with the sideband frequencies causes diffusion and additional particle loss. In the next section we discuss theses effects and show how the time dependent optimization reduces them.

\section{Sideband Instability}

\begin{figure*}
\includegraphics[scale=0.45]{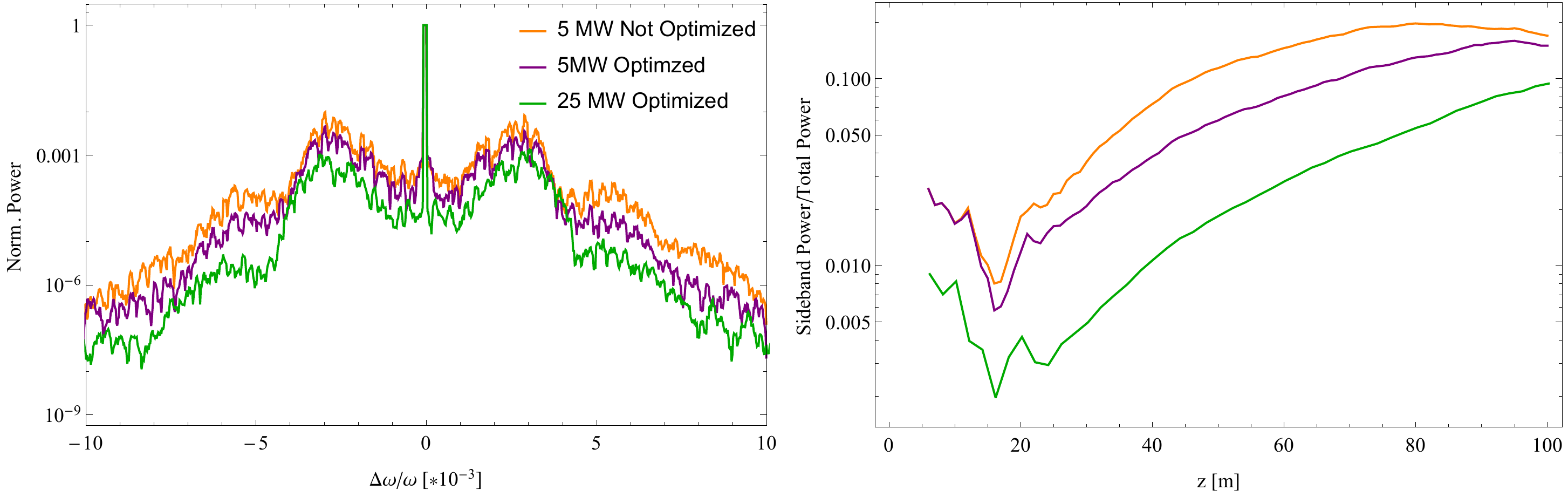}
\caption{(left) Spectrum at half the undulator length and fractional sideband power (right) for the time independent optimized case and fully time dependent optimized for 5 MW and 25 	MW input seed power and 1.5 MeV energy spread. Sideband power grows faster in the time independent optimized case, leading to particle detrapping and early saturation of the tapered FEL. }
\end{figure*}

The mechanism of sideband generation and amplification in free electron lasers can be summarized as follows \cite{krollbook}. Firstly, sidebands are generated due to amplitude and phase modulations of the electric field, due to the trapped particles undergoing synchrotron oscillations as they pass through the undulator. Using Maxwell's equations in the 1-D slowly varying envelope approximation we can write the evolution of the electric field amplitude and phase \cite{krollbook}:

\begin{eqnarray}
a_s'=\frac{\omega_p^2}{2\omega_s c}a_w \left\langle\frac{\sin \Psi}{\gamma}\right\rangle \\
\delta k_s=\frac{\omega_p^2}{2\omega_s c} \frac{a_w}{a_s} \left\langle\frac{\cos \Psi}{\gamma}\right\rangle 
\end{eqnarray}

where $a_s$ is the dimensionless vector potential for the electric field, $\omega_p$ is the electron beam plasma frequency and $\Psi$ is the ponderomotive phase. It is clear from these that as the electrons oscillate in the longitudinal phase space $(\Psi,\gamma)$ the gain and the phase shift of the radiation field will be different at different locations in the undulator and, due to shot noise in the electron beam, at different locations along the bunch. This results in a temporal modulation of the radiation field giving rise to sidebands displaced from the central wavelength by a quantity proportional to the synchrotron period:
\begin{equation}
\lambda_{s'}\approx \lambda_s\left[1\pm \frac{\lambda_w}{L_{sy}}\right]=\lambda_s\left[1\pm \left(\frac{a_w a_s}{1+a_w^2}\right)^{1/2}\right]
\end{equation}

where $L_{sy}$ is the synchrotron period. Once the sidebands are generated, the electron oscillations are driven by a multiple frequency ponderomotive potential, therefore the equations of motion and Maxwell's equations for the electric field, must be modified accordingly. An analysis of the simplest two frequency model shows that the coupled beam-radiation system is unstable and that the sideband amplitude will grow from noise for any realistic electron distribution \cite{krollbook} \cite{RiyopoulosTangSidebands}. When the strength of the sidebands exceeds a critical level, electron motion becomes chaotic leading to severe particle detrapping and a loss of amplification of the FEL signal \cite{RiyopoulosTangChaos}. Thus, as has been discussed by previous authors, suppressing the sideband instability is the key issue for tapered FEL designs \cite{KMR}, particularly those which are multiple synchrotron periods in length \cite{Quimby}. 

As is shown in Fig. 4 the time dependent optimized taper profile reduces sideband amplitude growth. This results in a reduction in particle loss and a delayed taper saturation, both evidenced in the increased bunching factor and output power in Fig. 2. In the simple case of constant sideband and carrier amplitude the diffusion coefficient caused by sideband excitations is proportional to the ratio of the power in the sidebands to the power in the FEL signal $D\propto C P_{s'}/P_s$ with the coefficient C depending on the type of sideband spectrum \cite{RiyopoulosTangChaos}. As is also shown in Fig. 4 this is reduced in the time dependent optimized case. The peak power improves by 1 TW between the time dependent optimized and un-optimized cases, an overall improvement of 2 $\%$. Despite the dedicated time dependent optimization we do not recover the single bucket extraction efficiency unlike results previously reported in Ref. \cite{HafiziTangTaperedSidebands}.  

\subsection{Effect of the seed power}

Increasing the ratio of input seed power to equivalent shot noise power in the electron beam minimizes the impact of sideband growth as shown in Fig. 5. We have analyzed the impact of this effect numerically by performing the same time dependent optimization described above with a 25 MW input seed assuming the same initial energy spread of $\sigma_E=$1.5 MeV for comparison purposes. While time independent tapering optimizations produce a peak power of 7.7 TW very similar to the 5MW seed case, time depedent optimizations have a much better performance, with a final output power of 6.3 TW. This is shown in Fig. 6 where we can see the peak radiation power still growing after 100 m and the bunching factor decaying slowly in the tapered section of the undulator. The mitigation of sideband induced detrapping in this case is also evidenced by the fractional sideband power which remains below 10 $\%$ throughout the undultaor (see Fig. 4).

\begin{figure}
\includegraphics[scale=0.4]{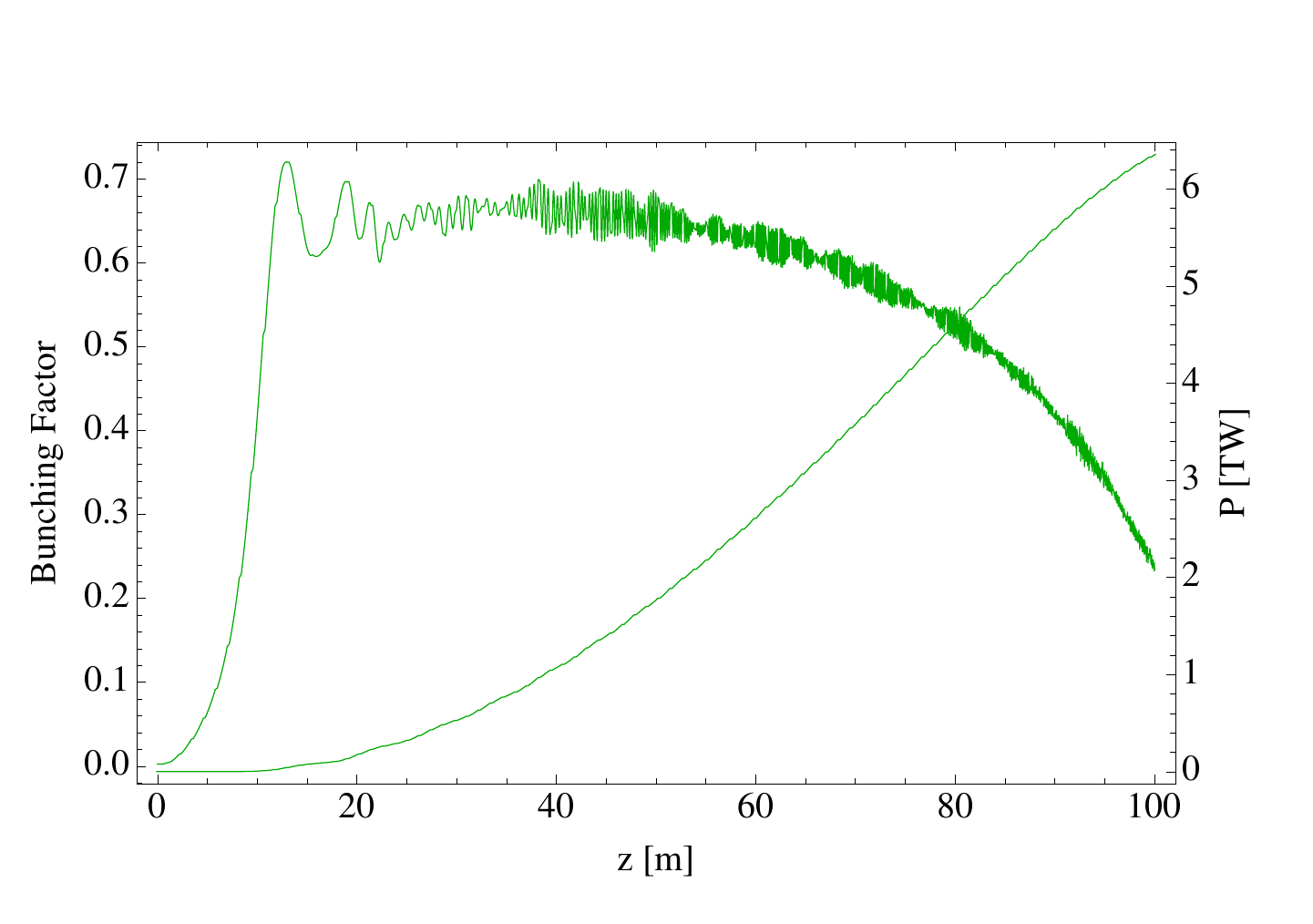}
\caption{Time dependent optimized power and taper profile for a 25MW seed and 1.5 MeV input energy spread. The bunching factor decays slowly due to reduced sensitivity to sideband induced detrapping and the radiation power does not saturate.}
\end{figure}

\subsection{Sideband suppression}

In order to reach the single bucket extraction efficiency we identify three separate solutions for further suppressing sideband growth which are currently being investigating. Firstly, as was pointed out originally in Ref. \cite{krollbook} and demonstrated numerically by \cite{GoldsteinColson}, increasing the electron beam energy spread in the last region of the undulator where the sideband amplitude is larger can also reduce the sideband growth. The energy spread can be introduced by means of a magnetic delay line and the interaction of the beam with an external laser inside a short, few gain length undulator around the location of exponential saturation following the self seeding monochromator. Depending on the flexibility of the undulator design one may also obtain this additional energy spread by detuning a number of undulators around the exponential saturation location and allowing the beam to radiate spontaneously outside the FEL gain bandwidth \cite{SaldinSpontaneous}.

Furthermore, a wavelength filter with corresponding delay line for the electrons could be placed in the tapered section in order to select a narrow bandwidth signal \cite{Quimby}. This should be done before the sideband power reaches the stochasticity threshold beyond which the trapping efficiency is seriously degraded. Filtering out the sideband frequencies will result in a reduced bucket height and some initial particle detrapping. This should be compensated by a reduction in sideband induced detrapping between the filter and the end of the undulator, thus providing a lower limit to the bandwidth of the filter. Lastly, we note that the time-independent efficiency can be achieved in a fully time dependent simulation by artificially removing the shot-noise from the electron beam distribution. While this is not entirely achievable in practice, the shot noise suppression method discussed in Ref. \cite{Ratnershotnoise} could be applied if the scheme can extended to hard X-ray wavelengths.

\section{\label{sec:level2}Conclusion}

In this paper we perform the first comparison of time independent and time dependent tapering optimization for a high efficiency seeded hard X-ray FEL. The comparison is done for an undulator design optimized to achieve TW peak powers in the shortest possible distance: helical, superconducting and with built-in focusing. We demonstrate that the taper profile yielding the maximum power in time independent optimizations does not correspond to the optimal solution when time dependent effects are included in the simulation. By performing time dependent scans of the taper order and the taper strength we show that the maximum output power in time dependent mode is achieved with a lower taper order compared to the time independent case. The difference is due to the increased sensitivity to particle detrapping in the time dependent case, mitigated by a slower taper profile in the later stages of the undulator where time dependent effects are more important. For an input energy spread of 3.1 MeV the final output power increases from 2.7 TW with the time independent taper profile to 3.7 TW with the profile obtained from dedicated time dependent scans. 

We have also discussed the importance of the trade-off between energy spread and seed power at the entrance of the tapered undulator section. We show that using a ``fresh bunch" with input energy spread of 1.5 MeV determined only by the linac we can decrease particle detrapping, maintain a larger bunching factor and improve the over-all performance. For the same seed power of 5 MW the maximum output power is 4.7 TW after the dedicated time dependent optimization. In a double-bunch system the input seed power can be larger without affecting the input energy spread. We have studied an optimal case with a 25 MW seed and 1.5 MeV energy spread and found that the output power reaches 6.3 TW at the end of the undulator, a 12 $\%$ efficiency which approaches the time independent result of 7.7 TW.  

We identify the sideband instability as the fundamental time dependent effect which is not taken into account in time independent optimizations and limits the extraction efficiency by causing particle detrapping. Analyzing the fraction of energy in the sidebands in the $\sigma_E=$1.5 MeV case with a 5 MW seed, we show that the fraction of energy deposited in the sidebands is below 10 $\%$ for 70 m in the time dependent optimized taper profile while it exceeds 10 $\%$ after 40 m in the the time independent case reaching $14 \%$ towards the end of the undulator. 

While extending the simulation method of Ref. \cite{PhysRevSTAB.Y.Jiao} to include time dependent effects significantly improves the performance of tapered X-FELs the current procedure is both time consuming and simulation intensive. The form of the taper profile $a_w(z)$ needs a more complicated functional dependence to optimize trapping in the early stages and reduce sideband growth in the later stages where time dependent effects play a more important role. With the enhanced understanding gained of the critical parameters limiting performance, such as the growth of the sideband instability, an improved algorithm can be developed which acts to directly suppress these effects. Such a scheme will be developed in future work. 

\section*{ACKNOWLEDGMENT}
The authors would like to thank J. Duris, P. Musumeci, G. Marcus and A. Marinelli for useful discussions. We also acknowledge W. Fawley for sharing his expertise particularly with regards to the sideband instability.  This work was supported by U.S. D.O.E. under Grant No. DE-SC0009983.

\bibliographystyle{ieeetr}
\bibliography{prstabreferences.bib}

\end{document}